\documentclass[aps,pra,letterpaper,twocolumn,showpacs,amsmath,amssymb,superscriptaddress]{revtex4-1}
\usepackage{graphicx}
\usepackage{setspace}
\usepackage{dcolumn}
\usepackage{mathrsfs}
\usepackage{bm}
\usepackage{longtable}
\usepackage{xcolor}
\usepackage{braket}
\usepackage{hyperref} % For hyperlinks in the PDF
\hypersetup{
	colorlinks=true,
	linkcolor=blue,
	citecolor=red,
	urlcolor=blue
}
\newcommand{\broad}{$5\mathrm{s^2}\, ^1\textrm{S}_0 - 5\mathrm{s}5\mathrm{p}\, ^1\textrm{P}_1$\,}
\newcommand{\narrow}{$5\mathrm{s^2}\, ^1\textrm{S}_0 - 5\mathrm{s}5\mathrm{p}\, ^3\textrm{P}_1$\,}

\begin{document}
\preprint{}

\title{Observation of photon recoil effects in single-beam absorption spectroscopy with an ultracold strontium gas}

\author{Fachao Hu}
\altaffiliation{These authors contributed equally to this work.}
\author{Canzhu Tan}
\altaffiliation{These authors contributed equally to this work.}
\affiliation{Hefei National Laboratory for Physical Sciences at the Microscale, University of Science and Technology of China, Hefei, Auhui 230026, China}
\affiliation{CAS Center For Excellence in Quantum Information and Quantum Physics, University of Science and Technology of China, Hefei, Anhui 230026, China}

\author{Yuhai Jiang}
\email{jiangyh@sari.ac.cn}
\affiliation{Shanghai Advanced Research Institute, Chinese Academy of Sciences, Shanghai 201210, China}
\affiliation{CAS Center For Excellence in Quantum Information and Quantum Physics, University of Science and Technology of China, Hefei, Anhui 230026, China}

\author{Matthias Weidem\"uller}
\email{weidemueller@uni-heidelberg.de}
\affiliation{Hefei National Laboratory for Physical Sciences at the Microscale, University of Science and Technology of China, Hefei, Auhui 230026, China}
\affiliation{CAS Center For Excellence in Quantum Information and Quantum Physics, University of Science and Technology of China, Hefei, Anhui 230026, China}
\affiliation{Physikalisches Institut, Universit\"at Heidelberg, Im Neuenheimer Feld 226, 69120 Heidelberg, Germany}

\author{Bing Zhu}
\email{bzhu@physi.uni-heidelberg.de}
\affiliation{Physikalisches Institut, Universit\"at Heidelberg, Im Neuenheimer Feld 226, 69120 Heidelberg, Germany}
\affiliation{CAS Center For Excellence in Quantum Information and Quantum Physics, University of Science and Technology of China, Hefei, Anhui 230026, China}

\date{\today}

\begin{abstract}
	We report on observing photon recoil effects in the absorption of a single monochromatic light at 689~nm through an ultracold $^{88}$Sr gas, where the recoil frequency is comparable to natural linewidth of the narrow-line transition \narrow in strontium. In the regime of high-saturation, the absorption profile becomes asymmetric due to the photon-recoil shift, which is of the same order as the natural linewidth. The lineshape is described by an extension of the optical Bloch equations including the momentum transfers to atoms during emission and absorption of photons. Our work reveals the photon recoil effects in a simplest single-beam absorption setting, which is of significant relevance to other applications such as saturation spectroscopy, Ramsey interferometry, and absorption imaging.
\end{abstract}

\maketitle

\section{Introduction}
When emitting or absorbing a photon, the motional degree of freedom of an atom is modified due to the so-called photon recoil effect, which is well-known and has triggered major breakthroughs in modern atomic physics through the developments of laser cooling and trapping techniques \cite{Cohen1990}. This effect also arises special interests in atomic spectroscopy \cite{Kolchenko1969} and interferometry \cite{Weiss1993}, single-photon scattering on a motional scatter \cite{Li2013}, and collective and cooperative light scattering in atom arrays for quantum information or control of light \cite{Robicheaux2019, Shahmoon2019}.

The study of photon recoil effects on atomic spectra dates back to 1970s in the context of laser technology and precision measurements \cite{Hall1978}. Three different frequencies are important when considering such an effect for an ensemble of two-level atoms irradiated by a monochromatic light, namely the natural linewidth of the atomic transition $\Gamma$, the recoil frequency $\gamma=\hbar k^2/2m$, and the Doppler shift $kv_0$. Here $\hbar$ is the reduced Planck constant, $k=2\pi/\lambda$ is the light wavenumber ($\lambda$ the light wavelength), $m$ is the atomic mass, and $v_0$ is the most probable velocity in the atomic ensemble. In case that $\Gamma\gg\gamma, kv_0$, the photon recoil can manifest itself in the effects of recoil-induced resonances ($kv_0\gg\gamma$) \cite{Guo1992, Courtois1994} and the collective atomic recoil laser ($kv_0\lesssim\gamma$) \cite{Bonifacio1994, Slama2007}. When $\gamma$ becomes comparable to or even larger than $\Gamma$, the photon recoil effects can cause a splitting in saturation spectroscopy, which was first predicted more than 60 years ago \cite{Kolchenko1969} and demonstrated in various experiments \cite{Hall1976, Riehle1988, Bagayev1989, Oates2005}. The cumulative recoils lead to spontaneous scattering force, as studied also by saturation spectroscopy in Refs. \cite{Grimm1988, Grimm1989, Grimm1989a, Grimm1990, Minardi1999, Artoni2000, Zheng2019}, although all the studies were performed with at least two laser beams for heated vapor gases or fast atomic beams and $\Gamma$ is still considerably larger than $\gamma$. 

In this work, we show the photon-recoil-induced asymmetry and shift in the absorption profile on the narrow-line transition \narrow at 689 nm with an ultracold $^{88}$Sr atomic cloud, where $\Gamma/2\pi=7.5$~kHz, $kv_0/2\pi\sim24$~kHz, and $\gamma/2\pi=4.8$~kHz. By monitoring the transmission of a single monochromatic beam through an ultracold dilute strontium gas (the atomic density $n\sim8.9\times10^{10}$)~cm$^{-3}$, we observe asymmetrical and shifted spectra depending nonlinearly on the light intensity. The simple experimental configuration studied here allows us to compare the measurements quantitatively to a theoretical simulation based on optical Bloch equations (OBEs) involving the momentum transfers during the  photon absorption and emission processes. Although the photon recoil effect was investigated extensively for $\gamma\lesssim\Gamma$ starting from more than four decades ago \cite{Hall1976, Riehle1988, Bagayev1989, Oates2005}, all the studies involved complex beam configurations like saturation or Ramsey spectroscopic and it has been elusive for observing corresponding phenomena using a single-beam setup. This work will be relevant to high-resolution optical spectroscopy with strontium or other alkali-earth (like) systems \cite{Christensen2015, Westergaard2015, Hu2017, Hu2019a, Rudolph2020}.

The article is organized as follows: We show our experimental setup in Sec. \ref{sec:setup}. The observation of photon recoil effects in the absorption spectra are described in Sec. \ref{sec:measurement} and the theoretical simulation and comparison to experiments are discussed in Sec. \ref{sec:sim}. Sec. \ref{sec:conclusion} concludes the paper.

\section{Experimental setup} \label{sec:setup}

\begin{figure}[t]
	\centering
	\includegraphics[width=0.42\textwidth]{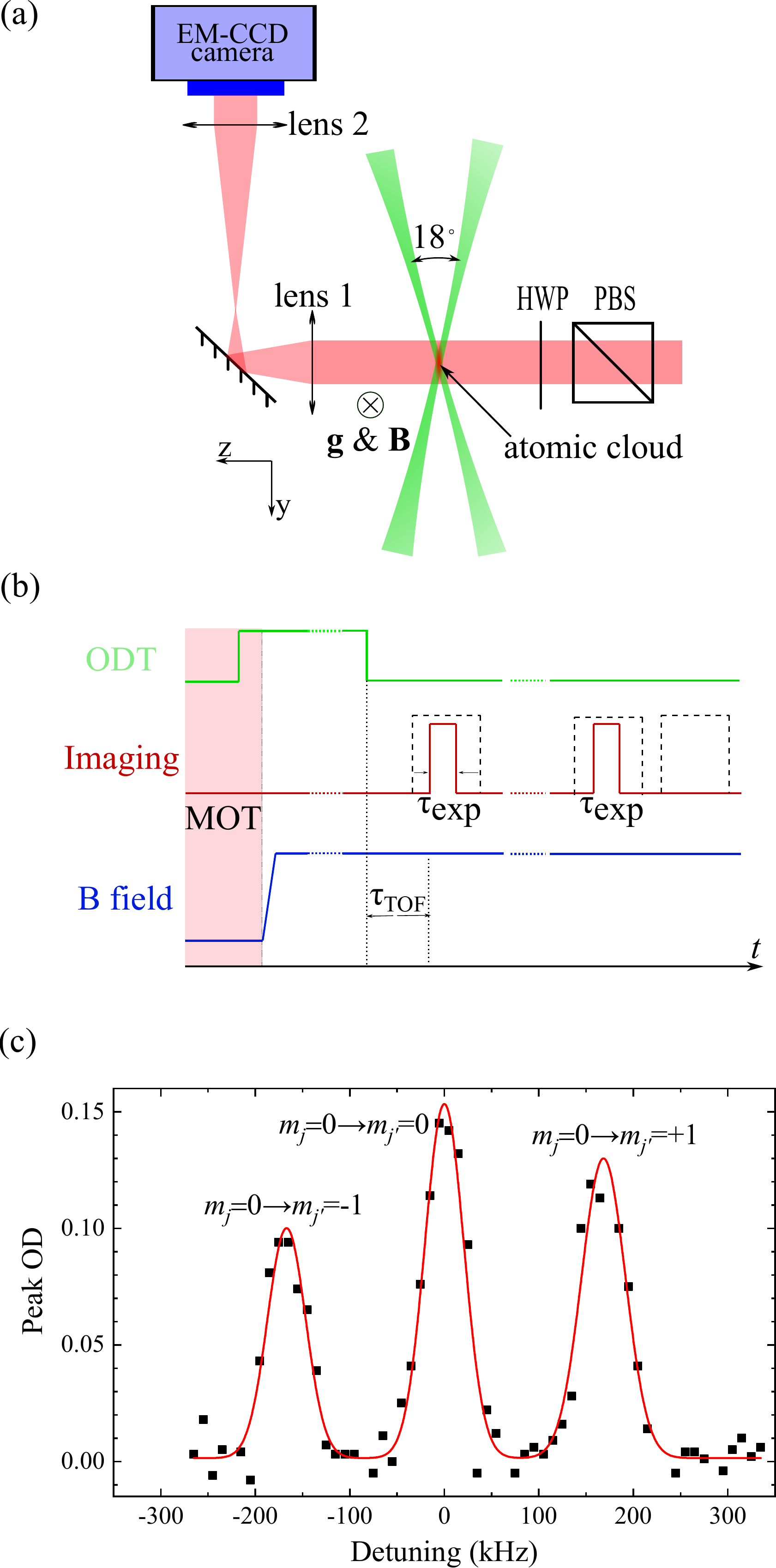}
	\caption{(a) Schematic of the top view of experimental setup. HWP: half wave-plate; PBS: polarizing beam-splitter. \textbf{g} and \textbf{B} represent the gravity and magnetic field, respectively. See text for more details. (b) Time sequence for absorption spectroscopy. See text for explanations of $\tau_{\textrm{TOF}}$ and $\tau_{\textrm{exp}}$. (c) Absorption spectrum showing all three Zeeman sublevels of $^3\textrm{P}_1$ state when the probe light polarization is tuned to about 45$^{\circ}$ angled to the residual magnetic field. Black points are the measured peak OD, and the red curve is the fit to a multi-peak Gaussian function. The obtained Zeeman splitting is 167.7(1.2)~kHz, corresponding to a magnetic field of 79.9(6)~mG.}
	\label{fig1}
\end{figure}

Fig. \hyperref[fig1]{1(a)} shows the experimental setup. The $^{88}$Sr atoms are first loaded into a two-stage magneto-optical trap (MOT) for the laser cooling and trapping \cite{Nosske2017, Qiao2019}, operated on the broad \broad and narrow \narrow transitions, respectively. Typically we produce an atomic cloud of $10^6$ atoms with a density of about $10^{10}$~cm$^{-3}$ and a temperature around 1~$\mu$K in the MOT. A cigar-shaped optical dipole trap (ODT) formed by two horizontally propagating beams at the wavelength of 532~nm, is simultaneously switched on during the narrow-line MOT \cite{Qiao2019}. The two ODT beams both have a waist of about 60~$\mu$m and cross at an angle of 18$^{\circ}$. Holding atoms in the ODT for 200~ms to reach equilibrium after switching off the MOT, we obtain about $(0.5\cdots 5)\times10^5$ atoms at a temperature of $0.7 \cdots 6$~$\mu$K depending on the ODT power. At a power of 0.6~W for each beam the trap depth of the ODT is about $6\mu$K and the trap frequencies are $2\pi\times$(217, 34, 217)~Hz along the $x$, $y$, and $z$ directions [see Fig. \hyperref[fig1]{1(a)}], respectively, resulting in cloud radii of (27, 69, 27)~$\mu$m and a peak density of $7\times10^{11}$cm$^{-3}$. The above-mentioned atom numbers, cloud sizes, and temperatures are measured using standard time-of-flight (TOF) imaging technique \cite{Ketterle1999} with the broad \broad transition ($\Gamma/2\pi\approx32$~MHz). The lifetime of the atomic clouds in the ODT is about 2~s, limited by the collisions with background particles.

The probe light at 689~nm is delivered from a commercial tapered amplifier seeded by an external-cavity diode laser (Toptica TApro), used also for the narrow-line MOT cooling, which is frequency-stabilized to a passive ultra-low expansion cavity with a short-term noise of 1~kHz level and a long-term drift of 8~kHz/day \cite{Qiao2019}. As shown in Fig. \hyperref[fig1]{1(a)}, the probe beam propagates along the $z$ direction with a tunable linear polarization and has a $1/e^2$ diameter of 4.2~mm. The probe pulse length and intensity are controlled by an accousto-optic modulator (not shown in the figure). The intensity distribution of the transmitted probe light is mapped out after passing two achromatic lenses with focal lengths of +200~mm and +300~mm by an EM-CCD camera from Andor, resulting in a resolution of about 12~$\mu$m [see Fig. \hyperref[fig1]{1(a)}].

\begin{figure*}[ht]
	\centering
	\includegraphics[width=\textwidth]{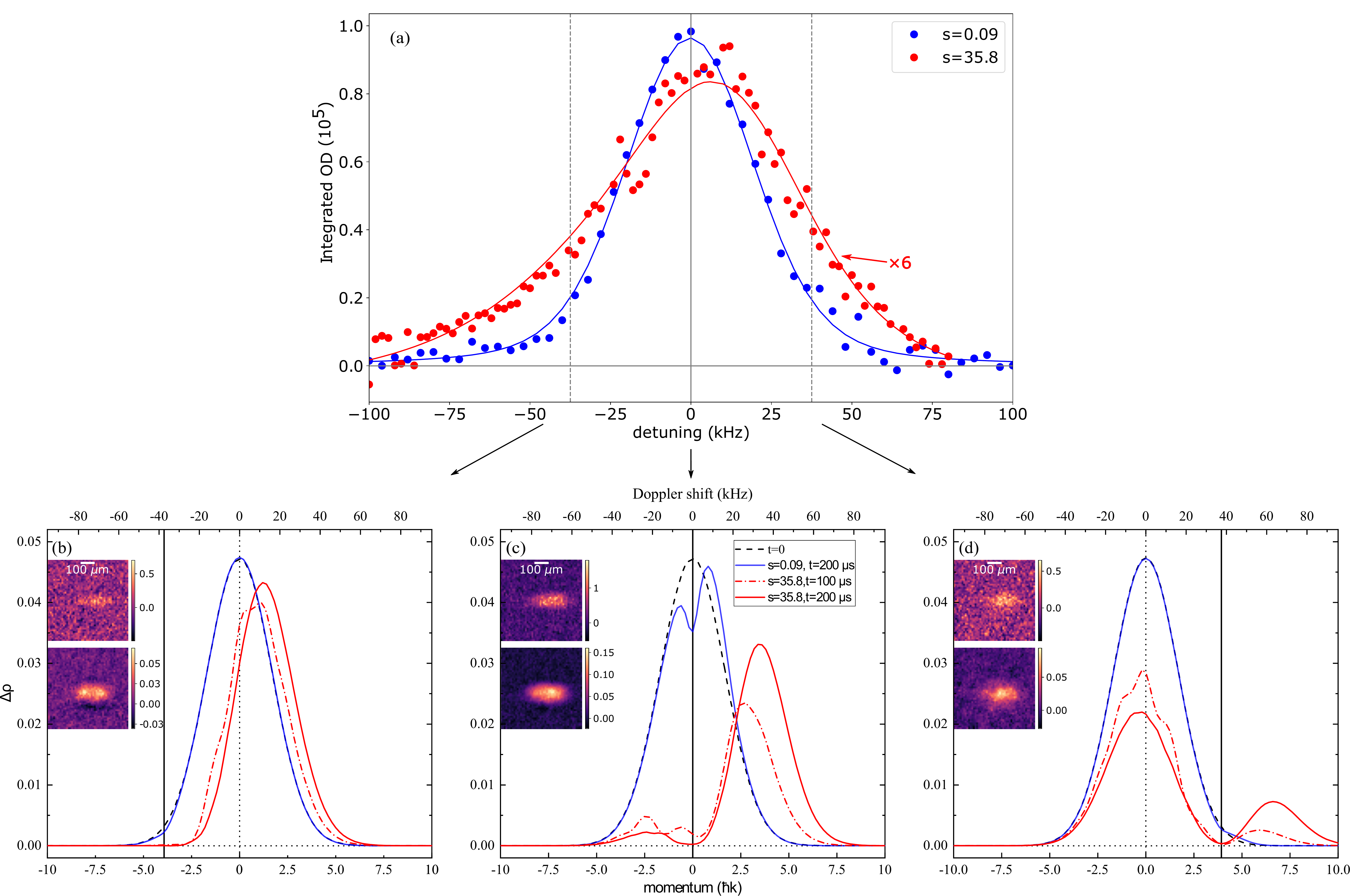}
	\caption{Photon recoil effects. (a) The measured absorption lineshapes at low (blue dots) and high (red dots) saturations. The data in the high-saturation case ($s=35.8$) showing asymmetric profile is fitted to the numerical solution of Eq. \eqref{eq:obe} (red curve) and magnified by 6 times to have a better visualization. As a comparison, the low-saturation ($s=0.09$) data is symmetric and fits well to the Voigt profile (blue curve). (b) - (d), the population difference $\Delta\rho_\delta(p)$ obtained from the OBE solutions at there different detunings [$0,\pm5\Gamma$, as marked by the grey vertical lines in (a)] after an exposure time of 100~$\mu$s and 200~$\mu$s, respectively. As a reference, we also show the initial distribution at $\tau_{\textrm{exp}}=0$, which is the Maxwell-Boltzmann one determined by the cloud temperature. The black solid vertical lines mark the resonant momentum positions, where the probe detuning is compensated by the Doppler effect. The inset images show measurements of the two-dimensional OD distributions at low (upper) and high (lower) saturations for their respective detunings.}
	\label{fig:asym}
\end{figure*}

The experimental sequence is described in Fig. \hyperref[fig1]{1(b)}. The absorption spectroscopy is performed after rapidly switching off the ODT to avoid the differential AC Stark shifts on the energy levels. A quantization magnetic field along the vertical direction is imposed (rising time 2~ms) before the probe pulse to split the Zeeman sublevels of $^3\textrm{P}_1$, as seen in Fig. \hyperref[fig1]{1(b)}. After a given TOF time $\tau_{\textrm{TOF}}$, the atoms are shined by the probe light with an exposure time $\tau_{\textrm{exp}}=200$~$\mu$s. By tuning the $\tau_{\textrm{TOF}}$ we can tune the atomic density during the absorption, which is important to avoid the dispersive lensing effect (see Appendix \hyperref[ssec:lensing]{B}). $\tau_{\textrm{TOF}}$ is fixed to 3.1~ms if not specified explicitly. Two additional images with and without the probe light are taken after the first absorption pulse and the three images are then processed (see, e.g., \cite{Lewandowski2003}) to obtain the so-called two-dimensional optical density (OD) distribution of the atomic cloud [see the insets of Fig. \hyperref[fig:asym]{2(b,c,d)}], as done for the standard absorption imaging of ultracold atomic gases \cite{Ketterle1999}. 

By changing the linear probe polarization angle in the $x-y$ plane, all three Zeeman sublevels of the $^3\textrm{P}_1$ state are observable. An example is shown in Fig. \hyperref[fig1]{1(c)}. The peak OD is measured as a function of the probe detuning showing three peaks at a magnetic field of about 80~mG. The relative line strengths are determined by the polarization and the different coupling strengths of the three corresponding transitions (see Fig. \hyperref[fig1]{1(c)}). We have used this measurement to optimize the compensation of the background magnetic field to be better than 5~mG in our setup and to calibrate the quantization fields. For the absorption spectroscopy discussed below, we apply a field of 4~G to split the sublevels of $^3\textrm{P}_1$ further ($\sim8$~MHz) and the probe polarization is tuned parallel to the quantization axis, so that the system is subjected only to the closed $\pi$ transition ($m_{j}=0 \rightarrow m_{j'}=0$), which can be treated as a perfect two-level system.

\section{Measurements} \label{sec:measurement}

Thanks to the high sensitivity and large dynamical range of our imaging camera (Andor iXon 897) at 689~nm, we can study the absorption spectrum on the narrow-line transition with a saturation parameter $s$ ranging from 0.01 to more than 100. Here $s=I/I_s$ is the ratio between probe intensity $I$ and the saturation intensity $I_s=3~\mu$W/cm$^2$. Here we will focus on the narrow-line absorption spectrum at strong saturations with $s\gg 1$. Naturally, one would expect more photon recoil events for longer probe time $\tau_{\textrm{exp}}$. However, the signal-to-noise ratio (SNR) is already strongly suppressed for $\tau_{\textrm{exp}}>300\mu$s in our measurements (explained later). We chose $\tau_{\textrm{exp}}\ge100~\mu\mathrm{s}\sim5/\Gamma$ to ensure an observable recoil effect and $\tau_{\textrm{exp}}\le250\mu$s for a large enough SNR.

In Fig. \hyperref[fig:asym]{2(a)} we compare two absorption spectra with $\tau_{\textrm{exp}}=200~\mu$s at saturation parameters of $s=0.09$ (blue points) and $s=35.8$ (red points), respectively. Instead of the peak OD, the integrated one over the whole atomic cloud region is shown, which is to be compared with the OBE solutions including contributions from all atoms of different velocities (see Sec. \ref{sec:sim}). The low-saturation data shows a symmetric, zero-centered profile, which can be described very well by a Gaussian function with a width determined by the cloud temperature (see Appendix \hyperref[ssec:lensing]{A}). At high saturation, we observe a decrease of the integrated OD signal at all detunings due to the saturation effect (note that the data at higher saturation is magnified by a factor of 6 for a better view). Here the integrated OD refers to the sum of all the measured ODs in the atomic cloud region.  More importantly, the lineshape is asymmetric at the high saturation, namely the integrated OD approaches zero more slowly on the negative-detuning side than that on the positive one, and the absorption peak is shifted by a few kHz to the positive detuning. At high saturation, differences can already be seen in the OD images at the two detuning sides [see lower rows in the insets of Figs. \hyperref[fig:asym]{2(b-d)}], namely a wider spatial extension for the positive detuning than that for the negative one. In this series of experiments, the influence of the lensing effect on the OD measurement(see Appendix \hyperref[ssec:lensing]{B}) is negligible due to the low atomic densities involved here.

The observed asymmetry and peak shift can be interpreted qualitatively by considering the absorption process including the influence of the photon recoil. The photon recoil associated with each absorption-spontaneous emission cycle redistributes the momentum of atoms, which depends strongly on the light detuning \cite{Stenholm1978}. Consequently, an asymmetric lineshape and the shift of the maximum of absorption emerges when more and more photons are scattered due to the momentum redistribution in the atomic cloud. In order to resolve such effects, the Doppler width has to be comparable to the power-broadened line width. In the case of the strong saturation in Fig. \hyperref[fig:asym]{2(a)}, the power-broadened Lorentzian width $\Gamma \sqrt{1+s} \sim 45$ kHz is close to the Doppler one of $\sim 40$ kHz. In the following subsection a quantitative description is presented incorporating the photon-recoil effect in an OBE formalism. 

\section{Spectrum lineshape simulation} \label{sec:sim}
\begin{figure}[t]
  \centering
  \includegraphics[width=0.45\textwidth]{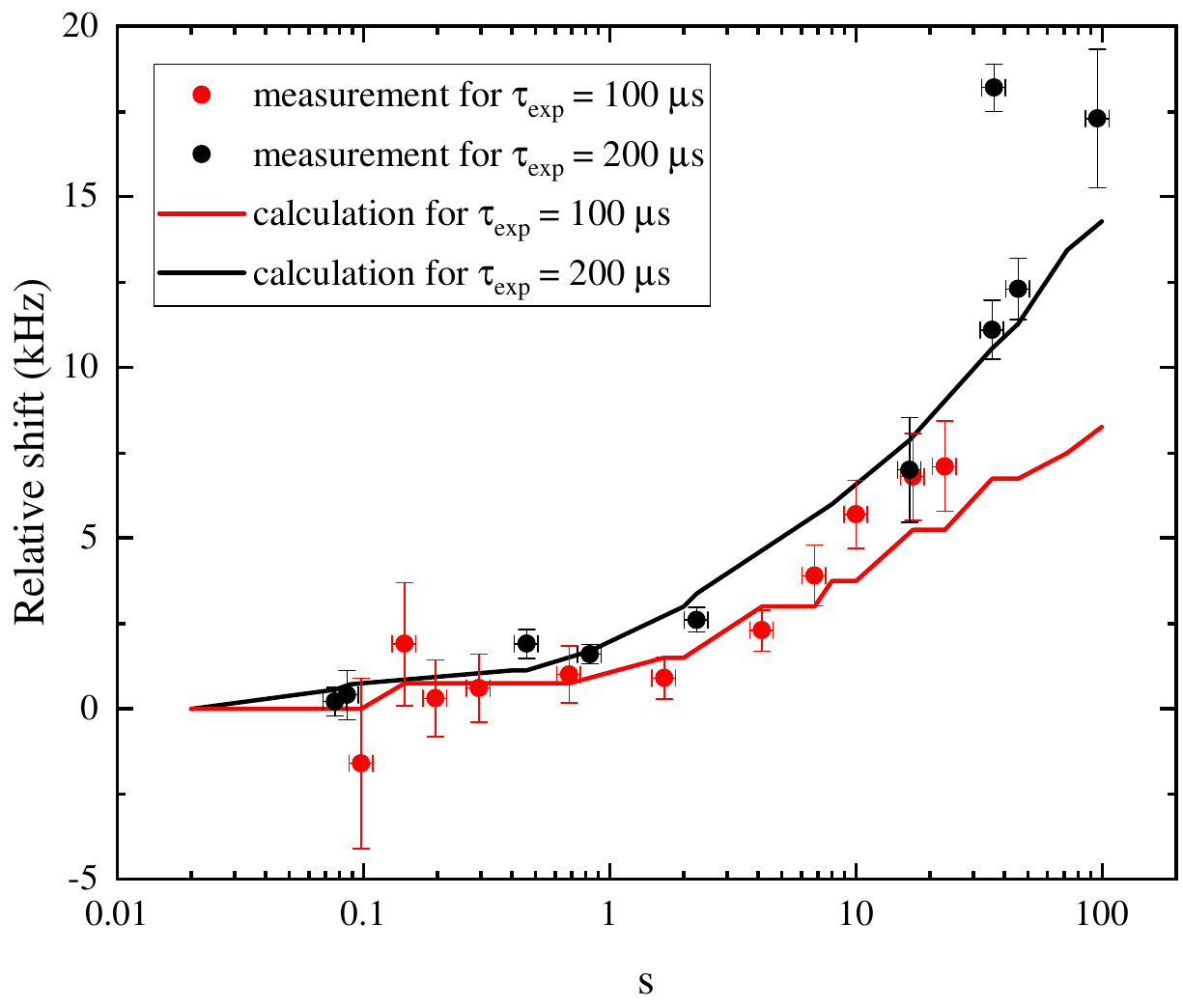}
  \caption{Absorption peak position shift. The relative position of the absorption peak at different saturation parameters $s$ are compared for imaging times of 100~$\mu$s (red circles and curve) and 200~$\mu$s (black circles and curve). The black and red curves are the calculated results without any free parameters, while the black and red circles are the fitted results from measurements by using the peak position and height as the two free fitting parameters. The error bars for the horizontal axis come mainly from uncertainties of measuring the probe power and those for the vertical axis are from fitting. See text for more discussions.}
  \label{fig:shift}
\end{figure}

We take an OBE formalism including the method of so-called 'momentum families' from Ref. \cite{Castin1989}, originally developed to understand laser cooling on a narrow-line transition. The model considers a two-level atom system with an initial Maxwell-Boltzmann thermal distribution, interacting with a single near-resonant monochromatic homogeneous probe beam. The state of an atom with momentum $\bm{p}$ is expressed in the form of $\{\ket{g,\bm{p}}, \ket{e,\bm{p}}\}$, where $\ket{g (e)}$ corresponds to the atomic ground (excited) state. 
The system Hamiltonian driven under a laser beam propagating along the $z$ axis is,
\begin{equation}
	\centering
	H_0 = \frac{\hat{\bm{p}}^2}{2m} + \hbar \omega_0 \ket{e}\bra{e} - \hat{\bm{D}}\cdot \hat{\bm{E}}
\end{equation}
where $\omega_0, \hat{\bm{D}}, \hat{\bm{E}}$ are the transition frequency, dipole moment operator, and laser electric field, respectively. In our case, only the $\pi$-transition branch $m_j = 0 \rightarrow m_{j'} = 0$ is considered, and only momentum along the light propagation axis $p=p_z$ is preserved, with the other two components $p_x,p_y$ traced over. The system Hamiltonian under the rotating-wave approximation becomes,
\begin{equation}
	\centering
	H_S = \frac{\hat{p}^2}{2m} - \hbar \delta\ket{e}\bra{e} +\frac{\hbar \Omega}{2}(e^{ikz}\ket{e}\bra{g} + \ket{g}\bra{e}e^{-ikz})
\end{equation}
where $\delta , \Omega$ are the bare detuning and Rabi frequency.

The evolution of states $\ket{g,p}, \ket{e,p+\hbar k}$ with any momentum p remains globally closed under $H_S$ when the spontaneous emission is not considered, for which reason the states $\ket{g,p}$, $\ket{e,p+\hbar k}$ are grouped as a family $\mathcal{F}(p)$. 
The system density matrix $\rho$ expanded in this basis is,
\begin{equation}
	\centering
	\begin{aligned}
		\rho_{gg}(p) &= \braket{g,p|\rho|g,p} \\
		\rho_{ee}(p) &= \braket{e,p+\hbar k|\rho|e,p + \hbar k} \\
		\rho_{ge}(p) &= \rho_{eg}^*(p)= \braket{g,p|\rho|e,p + \hbar k} \\
	\end{aligned}
	\, .
\end{equation}
The equations of evolution under $H_S$ together with the spontaneous emission processes are,
\begin{equation} \label{eq:obe}
	\centering
	\begin{aligned}
		\dot{\rho}_{gg}(p) &= \Gamma \bar{\pi}_e(p-\hbar k) - \frac{i\Omega}{2}(\rho_{eg}(p) - \rho_{ge}(p)) \, ,\\
		\dot{\rho}_{ee}(p) &= -\Gamma \bar{\pi}_e(p) + \frac{i\Omega}{2}(\rho_{eg}(p) - \rho_{ge}(p)) \, , \\
		\dot{\rho}_{ge}(p) &=\dot{\rho}_{eg}^*(p)\\
		& =-(i(\bar{\delta}-\frac{kp}{m})+\frac{\Gamma}{2})\rho_{ge}(p) + \frac{i\Omega}{2}(\rho_{gg}(p) - \rho_{ee}(p)) \, , \\
	\end{aligned}
\end{equation}
where $\bar{\delta}=\delta - \hbar k^2/(2m)$ and the term $\bar{\pi}_{e}$ represents the impact of spontaneous decay on the system evolution, defined as
\begin{equation} \label{eq:pi}
	\centering
	\begin{aligned}
		\bar{\pi}_{e}(p)=&\int\limits_{-\infty}^{+\infty}dp_x\int\limits_{-\infty}^{+\infty}dp_y\int\limits_{-\hbar k}^{+\hbar k}dp'\mathcal{N}(p')\\
		&\braket{e,p_x,p_y,p_z=p+p'|\rho|e,p_x,p_y,p_z=p+p'} \, .
	\end{aligned}
\end{equation} 
Here $\mathcal{N}(p')=\frac{3}{4\hbar k}(1-p'^2/\hbar^2k^2)$ results from the classical dipole radiation pattern \cite{Castin1989} of the $\pi$ transition. With all the atoms initially at the ground state $\ket{g}$ with a Maxwell-Boltzmann distribution of temperature $T$, we numerically integrate the equations \eqref{eq:obe} to get the system evolution. The solution of the off-diagonal elements $\rho_{eg}(p)$ results in the susceptibility $\chi(p)\propto n\rho_{eg}(p)$ with the atomic density $n$. The absorption profile is then calculated by tracing the imaginary part of the susceptibility over all momenta, i.e. $\sum_{p} \mathrm{Im}\chi(p)$, and then integrating over the interaction duration.

For the solid curves in Figs. \hyperref[fig:asym]{2(a)}, we fit the experimental data to the calculated profiles with the maximum integrated OD and the peak position as the only free parameters. Both the lineshape asymmetry and the shift of the absorption peak at high saturation can be reproduced very well by Eq. \eqref{eq:obe} including the momentum transfer due to the photon-scattering events. While the model predicts a significant shift of the absorption peak, its position is still used as a free parameter in the fits to account for the possible deviation between the measurements and the calculations, as discussed in more details later in Fig. \ref{fig:shift}.

One can gain further insight into the observed photon-recoil effects by considering the detuning-dependent distribution of the susceptibility in the momentum space, the imaginary part of which accounts for the absorption and reads $\mathrm{Im} \chi_\delta(p)\propto \frac{\Delta\rho_\delta (p)}{\Gamma/\Omega+(\delta/\Omega-kp/m\Omega)^2}$ in the quasi-steady solution of Eq. \eqref{eq:obe}. Here $\Delta \rho_\delta ( p)=\rho_{gg}(p) - \rho_{ee}(p)$ represents the population difference between the ground and excited states at a detuning $\delta$ for atoms with a momentum between $p$ and $p+dp$ such that $\int[\rho_{gg}(p) + \rho_{ee}(p)]dp=1$. If the photon recoils can be neglected, the integration of $\mathrm{Im} \chi(\delta, p)$ over the momentum space results in a symmetric absorption profile with respect to the detuning $\delta$. At all detunings the initial distribution is Gaussian determined by the gas temperature. In general, the momentum transferred to atoms during the photon scattering would reduce the absorption due to the parabolic dependence on $p$ in the denominator of $\mathrm{Im} \chi_\delta(p)$, and hence reducing the SNR for stronger recoils.

We show from Fig. \hyperref[fig:asym]{2(b)} to \hyperref[fig:asym]{2(d)} the calculated distribution of the population difference $\Delta \rho_{-5\Gamma, 0, + 5\Gamma} (p)$ in the momentum space at two saturation parameters of $s\approx0.09$ (blue curves) and $s=35.8$ (red curves) after 200-$\mu$s atom-light interaction time ($\tau_{\textrm{exp}}\sim10/\Gamma$ to ensure a quasi-steady state condition), respectively. At the low saturation ($s\approx0.09$), this distribution at all three $\delta$ is only slightly modified if compared to the initial Maxwell-Boltzmann distribution (black dot-dashed lines), remaining almost Gaussian even after long interaction time. This behaviour agrees with a weak photon-recoil effect and results in a lineshape nearly the Voigt one, as the blue curve seen in Fig. \hyperref[fig:asym]{2(a)}. When highly saturated  ($s=35.8$), however, the $\Delta\rho(p)$ distribution is strongly affected and we observe a significant depletion of atoms near the resonant momentum (marked by vertical dashed lines), where the Doppler shift compensates the bare imaging detuning. In Fig. \hyperref[fig:asym]{2(b)} with a detuning of $-5\Gamma$, the distribution maintains a Gaussian shape with the center shifted by $\sim1.2\hbar k$ after 200 $\mu$s. While at a detuning of $+5\Gamma$ in Fig. \hyperref[fig:asym]{2(d)}, two peaks appear on the opposite sides of the resonant momentum. Such a strong dependence on the detuning results from the fact that the probe light constantly transfers a positive momentum to the atomic ensemble and this transfer is highly momentum dependent for a narrow transition. We further note that to observe the asymmetric lineshape and the peak shift one would need also low sample temperatures to ensure the visibility of the momentum transfer effect.

The effects of the photon recoil can also be revealed by studying the time evolution of the momentum distribution $\Delta \rho_\delta(p)$. In Figs. \hyperref[fig:asym]{2(b-d)} the $\Delta \rho_{-5\Gamma,0,+5\Gamma}(p)$ at $s=35.8$ after 100-$\mu$s interaction (red dash-dot curves) are shown as a comparison to the 200-$\mu$s case. Small but clear differences of $\Delta \rho(p)$ are observed for all three detunings indicating that the momentum distribution undergoes some time evolution, which may result in a time-dependent absorption lineshape. This is actually demonstrated in Fig. \ref{fig:shift} by comparing the saturation-dependent shift of the absorption peak position for the 100- and 200-$\mu$s imaging durations. The peak position is shifted towards the positive detuning when increasing the imaging intensity and such a shift becomes larger in the case of a longer exposure, i.e. more photons are scattered. The solid curves represent the calculated results without any free parameters, while the solid dots are from fits with the peak position and height as the free fitting parameters [see Fig. \hyperref[fig:asym]{2(a)}]. Overall, the fitted shifts agree well with the calculations without free parameters, while deviations are seen for some points coming from fluctuations of experimental conditions like laser power and atom number, as well as the low SNR at strong saturations.

\section{Conclusion}
\label{sec:conclusion}
In conclusion, we have presented a simple experiment studying the photon recoil effects in the absorption spectrum of a narrow-line transition with an ultracold strontium gas, where the investigated parameter regime has not been experimentally studied before and a quantitative understanding of the observations is shown by OBEs using the "momentum family" method. Our study can be extended to high-density samples, in which the inter-particle distance is comparable to or smaller than the probe wavelength. Various collective and cooperative effects are predicted theoretically under these conditions \cite{Bienaime2012, Zhu2016, Kupriyanov2017, Bettles2020} and atomic motions are essential there. The narrow-line absorption studied here can also be employed as sensitive probes for detection of interaction effects in divalent systems, e.g. the spatial correlation due to Rydberg blockade \cite{Guenter2012}.

\section*{Acknowledgements}
We acknowledge C. Qiao, L. Couturier, I. Nosske and P. Chen for their contributions on setting up the experiment at the early stage of project. F.H acknowledges Yaxiong Liu for helpful discussions on numerical algorithms. We are supported by the Anhui Initiative in Quantum Information Technologies. Y.H.J. also acknowledges support from the National Natural Science Foundation under Grant No. 11827806.

\section*{Appendix} \label{appedix}

\subsection{Low-saturation absorption} \label{ssec:probe}

\begin{figure}[t]
	\centering
	\includegraphics[width=0.46\textwidth]{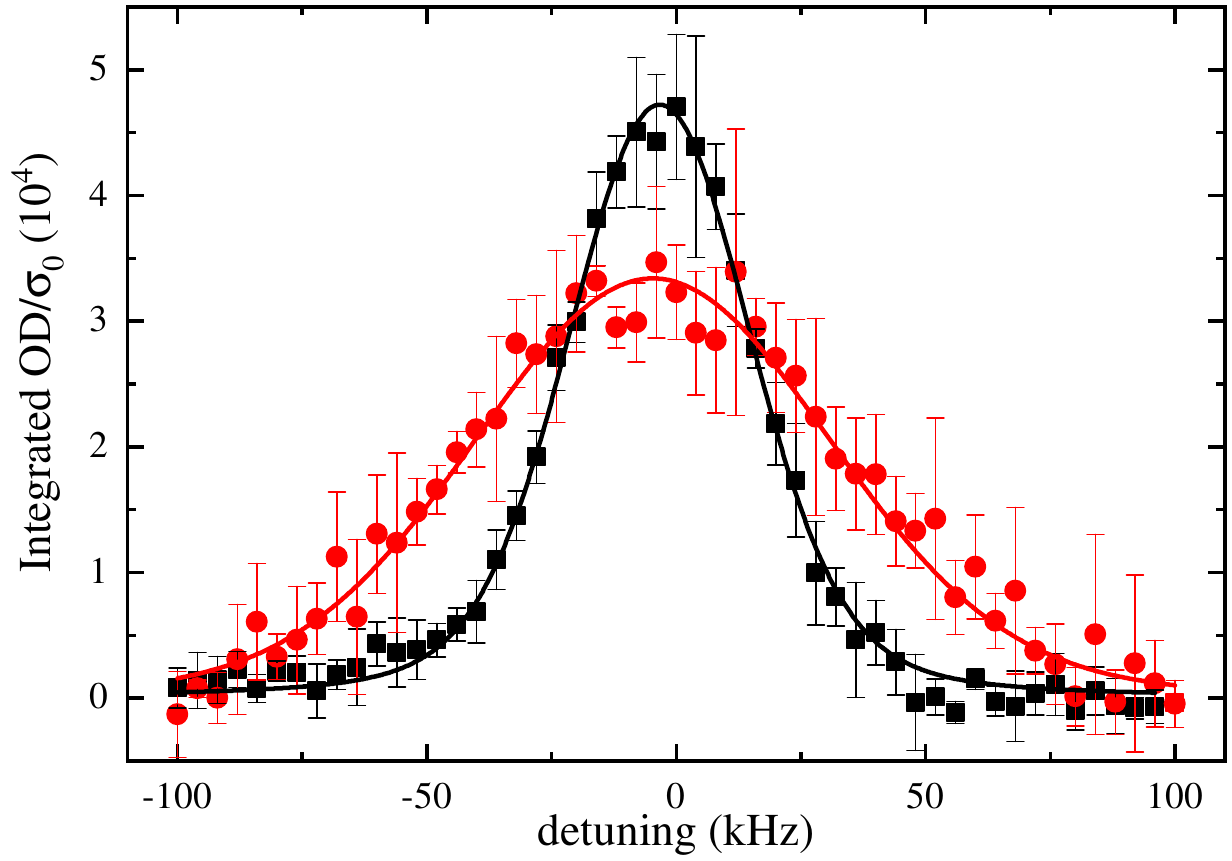}
	\caption{Low-saturation absorption spectra at temperatures of 1.3 $\mu K$ (black) and 5.7 $\mu K$ (red). The integrated absorption signal over the atomic cloud region is plotted as a function of the imaging detuning. The error bars are standard deviations out of five repeated measurements. The solid curves are fits to the Voigt profile. See text for more details.}
	\label{fig:Doppler}
\end{figure}

In Fig. \ref{fig:Doppler}, we show two measured absorption spectra at temperatures of 1.3~$\mu$K (black points) and 5.7~$\mu$K (red points) with a saturation parameter of $s = 0.1$. The TOF time $\tau_{\textrm{TOF}}$ (see Fig. \hyperref[fig1]{1(b)}) is chosen to be 3.1 ms to minimize the lensing effect (see Appendix \hyperref[ssec:lensing]{B}) as well as to keep large enough SNR in the OD images. The plotted signals in Fig. \ref{fig:Doppler} are the OD integrals over the whole atomic cloud region divided by the peak cross section $\sigma_0=3\lambda^2/2\pi$.

Symmetric lineshapes are observed in both cases and the linewidth increases with the increasing temperature. The spectra fit well to Voigt profiles with a fixed Lorentzian width of $v_L = 10.01$~kHz, resulted from the power broadening $\Gamma\sqrt{1+s}/2\pi$ and the detection bandwidth $0.9 /\tau_{\textrm{exp}}=4.5$~kHz due to the finite length of the square-shape imaging pulse (see Fig. \hyperref[fig:Doppler]{1(b)}), where $\Gamma/2\pi=7.5$~kHz is the natural linewidth. The FWHM Gaussian width $v_G$ obtained from the Voigt profile fitting agrees excellently with the Doppler width deduced from the cloud temperatures.

Other than revealing the temperature information, the atom number and atomic density can also be extracted from the narrow-linewidth absorption imaging in the low saturation regime. Broad transitions typically used in determining the atom number and atomic density have natural linewidths on the order of 10~MHz, much broader than the Doppler widths in the ultracold regime. The absorption cross-section in the broad-transition imaging can hence be regarded as temperature-independent. However, for the narrow transition with a natural linewidth smaller than the Doppler width ($\Gamma/2\pi v_G < 1$), the Doppler effect has to be considered when calculating the atom number \cite{Foot2004}. This is done by convolving the velocity-dependent Lorentzian absorption profile with the Maxwell-Boltzmann velocity distribution in the atomic sample. In general, the low-saturation ($s\ll1$) OD can be calculated as
\begin{equation} \label{eq:OD}
	\begin{aligned}
		OD(x,y) &= \sigma_0 n(x,y) \int_{-\infty}^{+\infty} \frac{1}{u\sqrt{\pi}}e^{-(v/u)^2} \frac{\Gamma^2/4}{(\delta-kv)^2 + \Gamma^2/4} dv \\
		& = \sigma_0 n(x,y) \frac{\alpha^2}{\sqrt{\pi}}\int_{-\infty}^{+\infty}\frac{e^{-(x'+\delta/ku)^2}}{x'^2+\alpha^2}dx' \\
	  & \overset{\delta=0}{\equiv} \sigma_0 n(x,y)\times C(\Gamma, v_G) \, ,
	\end{aligned}
\end{equation}
where $C(\Gamma, v_G) = \sqrt{\pi} \alpha e^{\alpha^2}\textrm{Erfc}(\alpha)$ is a coefficient with $\alpha=\sqrt{\ln2}\Gamma/2\pi v_G$ when the detuning $\delta$ is zero, and $n(x,y)$ is the atomic column density. Erfc(x) is the complementary error function. With the measured on-resonance OD and the temperature-dependent $v_G$ determined from the spectrum width, the atom number and atomic density can be obtained with Eq. \eqref{eq:OD}.

\subsection{The lensing effect}\label{ssec:lensing}

\begin{figure}[t]
	\centering
	\includegraphics[width=0.5\textwidth]{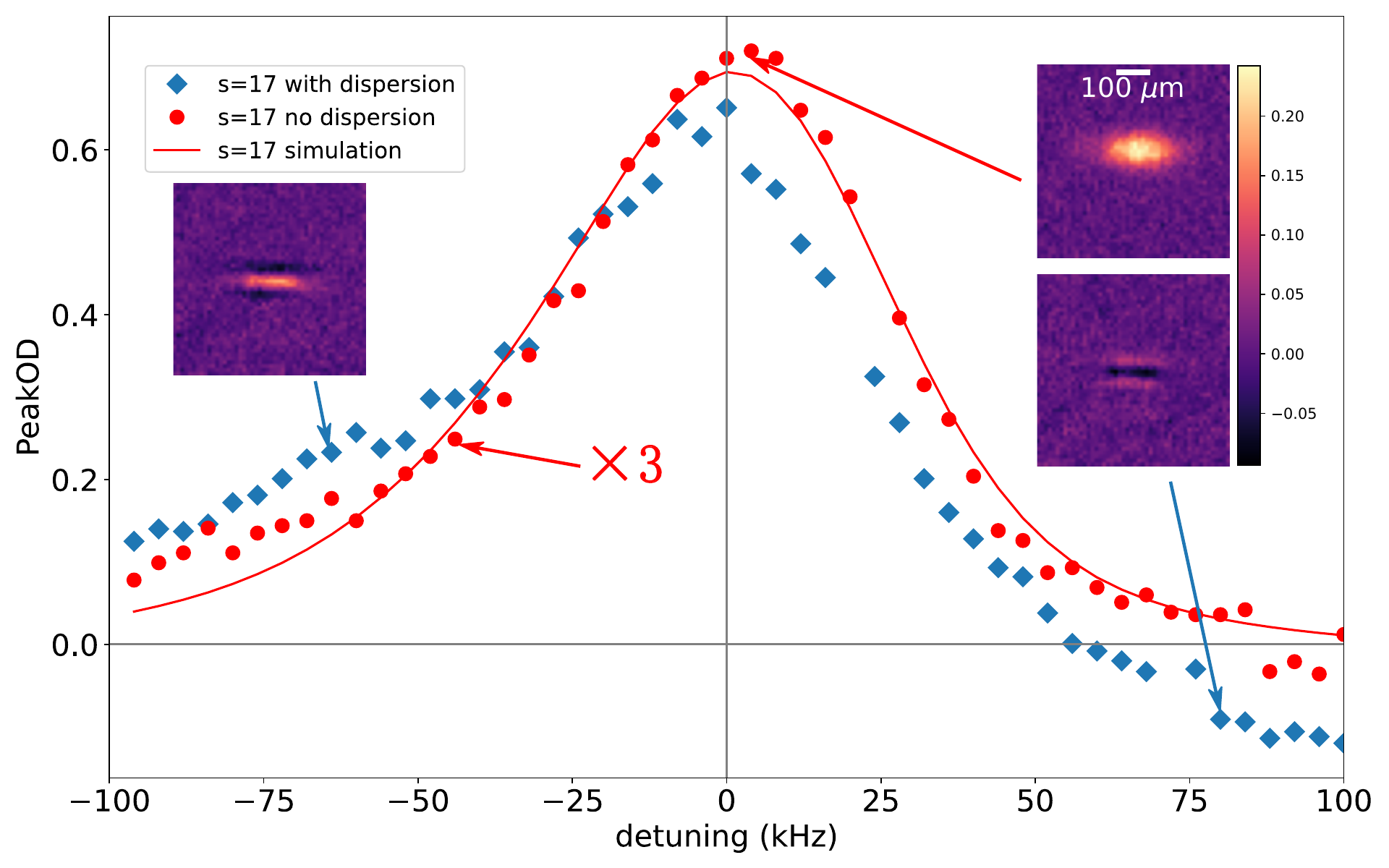}
	\caption{Absorption spectrum with $s=17$ at two different atomic densities of $8.9\times 10^{10}$~cm$^{-3}$ (red circles) and $2.8\times 10^{11}$~cm$^{-3}$ (blue diamonds). The red curve is a fit to the numerical solution of Eq. \eqref{eq:obe}. We obtain negative peak ODs at some large positive detunings. In the right inset, the lower OD image measured at a large positive detuning with the high atomic density has a dark hole instead of a bright peak in the cloud center, caused by the lensing effect. At the large negative detuning, the dark position appears at the edges of the cloud (left inset). As a comparison, we also show an example of the OD images for the low-density case with a normal Gaussian distribution in the upper panel of the inset. }
	\label{fig:dispersion}
\end{figure}

As shown in Fig. \ref{fig:dispersion}, we have also experimentally observed another phenomenon in the absorption spectrum at high atomic densities, the so-called lensing effect which is well known in standard absorption imaging. The absorption spectra at two different atomic densities, tuned by $\tau_{\textrm{TOF}}$, are compared at a saturation of $s=17$. For a better illustration of the lensing, we show the peak OD instead of the integrated one as a function of the detuning. In the low-density case ($\tau_{\textrm{TOF}}=3.1$~ms, $n\sim8.9\times 10^{10}$~cm$^{-3}$, red dots in the figure), we find a similar asymmetry as that in Fig. \hyperref[fig:asym]{2(a)} for the high saturation. With a 3-fold higher density ($\tau_{\textrm{TOF}}=1.1$~ms, $n\sim2.8\times 10^{11}$~cm$^{-3}$, blue diamonds in the figure), a negative peak OD is obtained from the two-dimension Gaussian fit at some large positive detunings. Checking the OD images there (one example shown as the right inset in Fig. \ref{fig:dispersion}), a dark hole instead of a bright peak is seen at the central region of the atomic cloud for the large positive detuning, while at the negative one a dark edge is observed. This phenomonon is related to the microscopic lensing effect studied in e.g. Refs. \cite{Labeyrie2003, Wang2004, Labeyrie2007, Roof2015, Han2015, Noaman2018, Gilbert2018}, where a spatial-dependent index of refraction leads to a focusing or defocusing effect on the imaging beam depending on the detuning.

The observed lensing effect here mainly stems from the density inhomogeneity as indicated by the density-dependence (see Fig. \ref{fig:dispersion}) and the fact that the imaging beam is much larger than the atomic cloud ($\sim200$ times). The lensing induced by such densitiy inhomogeneity was observed in both the weak- \cite{Roof2015} and strong-saturation \cite{Labeyrie2003, Wang2004, Labeyrie2007} regimes. The lensing effect shown in Fig. \ref{fig:dispersion} with strong saturation is also observable in the weak-probe case in our experiment. However, to quantitatively explain our observation, detailed calculations on the light propagation are needed like in Refs. \cite{Han2015, Gilbert2018}, even including the atom dipolar interactions or multiple scattering events (e.g. \cite{Bromley2016, Zhu2016, Chabe2014}), which is beyond the current scope.

\bibliography{mylibrary}

\end{document}